\documentclass[conference]{IEEEtran}
\IEEEoverridecommandlockouts

\usepackage{cite}
\usepackage{amsmath,amssymb,amsfonts}
\usepackage{algorithmic}
\usepackage{graphicx}
\usepackage{textcomp}
\PassOptionsToPackage{table,xcdraw}{xcolor}

\usepackage{float}
\usepackage{array}
\usepackage{caption}
\usepackage{doi}
\usepackage{placeins}  
\usepackage{listings} 
\usepackage{subcaption}

\usepackage[scaled=0.86]{helvet}
\usepackage{orcidlink}
\usepackage{xspace}

\newcommand{\AutoBIR}{\textsc{\textsf{AutoBIR}}\xspace}

\usepackage{mdframed} 
\definecolor{codegreen}{rgb}{0,0.6,0}
\definecolor{codegray}{rgb}{0.5,0.5,0.5}
\definecolor{codepurple}{rgb}{0.58,0,0.82}
\definecolor{backcolour}{rgb}{0.95,0.95,0.92}
\definecolor{specialwordcolor}{rgb}{1,0.5,0} 

\DeclareCaptionFormat{mylst}{\hrule#1#2#3}
\begin{document}

\title{Automating Business Intelligence Requirements with Generative AI and Semantic Search
\thanks{This research is supported by a MITACS-Accenture Accelerate grant titled \textit{Beyond Keywords: Semantic Search Framework for Data in Organization}.}
}

\author{\IEEEauthorblockN{Nimrod Busany~\orcidlink{0009-0000-2843-9515}}
\IEEEauthorblockN{Ethan Hadar}
\IEEEauthorblockN{Hananel Hadad~\orcidlink{0009-0009-3180-0020}}
\IEEEauthorblockN{Gil Rosenblum}
\IEEEauthorblockA{\textit{Accenture Labs} \\
 Herzelia, Israel \\
 \textsf{Nimrod.Busany, Ethan.Hadar, Hananel.Hadad,} \\
 \textsf{Gil.Rosenblum (@accenture.com)}
 }
 \and
 \IEEEauthorblockN{Zofia Maszlanka~\orcidlink{0009-0000-0243-2809}}
 \IEEEauthorblockA{\textit{Avanade} \\
 Krakow, Poland \\
 \textsf{Z.Maszlanka (@avanade.com)}
 }
 \and
 \IEEEauthorblockN{Okhaide Akhigbe~\orcidlink{0000-0003-1255-8135}}
 \IEEEauthorblockN{Daniel Amyot~\orcidlink{0000-0003-2414-1791}}
 \IEEEauthorblockA{\textit{EECS, University of Ottawa} \\
 Ottawa, Canada \\
 \textsf{Okhaide, DAmyot (@uottawa.ca)}
 }
}

\maketitle

\begin{abstract}
Eliciting requirements for Business Intelligence (BI) systems remains a significant challenge, particularly in changing business environments. This paper introduces a novel AI-driven system, called \AutoBIR, that leverages semantic search and Large Language Models (LLMs) to automate and accelerate the specification of BI requirements. The system facilitates intuitive interaction with stakeholders through a conversational interface, translating user inputs into prototype analytic code, descriptions, and data dependencies. Additionally, \AutoBIR produces detailed test-case reports, optionally enhanced with visual aids, streamlining the requirement elicitation process. By incorporating user feedback, the system refines BI reporting and system design, demonstrating practical applications for expediting data-driven decision-making. This paper explores the broader potential of generative AI in transforming BI development, illustrating its role in enhancing data engineering practice for large-scale, evolving systems. 
\end{abstract}

\begin{IEEEkeywords}
Business Intelligence, Generative AI, Large Language Models, AI-Driven Data Engineering, Requirements Automation, Semantic Search, Ontology-Based Query Generation, Text-to-SQL
\end{IEEEkeywords}

\section{Introduction}
In today’s data-driven landscape, organizations rely on Business Intelligence (BI) systems to extract insights from distributed data sources such as SAP software, Customer Relationship Management (CRM), and internal databases. As the business environment rapidly evolves,  BI systems must continually adapt to changing requirements, necessitating ongoing adjustment to analytics and data models to support informed decision-making processes \cite{akhigbe_framework_2014}. However, the process of eliciting and specifying these requirements remains a significant challenge, particularly as BI systems scale and requirements change dynamically.

Manually managing these evolving requirements is labor-intensive and error-prone, requiring significant coordination between data analysts, subject matter experts, and business stakeholders. Traditional methods of capturing BI requirements often lead to gaps between business needs and technical implementation~\cite{Weichbroth2024}. These inefficiencies can result in repeated design cycles, misalignment between available data and desired analytics, and disruptions in production environments. Furthermore, the increasing shift towards cloud adoption, data transformation, and technology migration~\cite{dehghani_data_2022} further complicates the process of ensuring that BI systems evolve without compromising business continuity. To address these challenges, automation tools are increasingly necessary to streamline the specification process, ensuring accuracy while reducing the need for repetitive manual interventions. 

Current approaches struggle with capturing the semantic nuances of data schemas and aligning business requirements with technical solutions, particularly in legacy systems. Related work by Yu et al.~\cite{Yu2013}, Teruel et al.~\cite{Teruel2014}, Horkoff et al.~\cite{horkoff2014strategic}, and Burnay et al.~\cite{burnay2016framework} have explored different usages of goal models for designing BI systems, but without automation. Lavalle et al.~\cite{Lavalle2019,Lavalle2021} have explored automated model-driven approaches, but with a focus on the selection of visualizations. To our knowledge, no approach is using recent advancements in Artificial Intelligence (AI) to automate the comprehensive specification of BI requirements.

We contribute a novel system that uses Generative AI, specifically Large Language Models (LLMs) and Semantic Search, to automate and expedite the specification of BI requirements. This no-code system allows users to interact with data through a conversational interface, translating natural language inquiries into executable queries that yield analytics specifications, test-case reports, data dependencies, and prototype analytic code. By reducing the time and effort required for BI system development while ensuring accuracy, this system represents a substantial advancement in AI-driven data engineering, particularly in the automation of requirements elicitation. Our architecture leverages semantic technologies to align user inputs with underlying data structures, ensuring accurate, contextually relevant queries. By incorporating user feedback, the system continuously refines its outputs, enhancing both the design and reporting capabilities of BI systems. Through the application of LLMs and Semantic Search, the system streamlines data management processes and opens new avenues for scalability and flexibility in BI system design.  

This paper explores the technical architecture of this system, its integration of AI and data discovery technologies, and its practical implications for advancing BI analytics development. It also contributes a discussion of the broader potential of Generative AI in transforming the landscape of data engineering, particularly in enabling more efficient, automated processes for managing and evolving large-scale BI systems.

This paper is structured to promote a comprehensive understanding of our novel system named \AutoBIR (Automated BI Requirements), and its development, application, and potential evolution. Section~\ref{sec:example} provides a practical example of user-system interaction. Section~\ref{sec:prem} reviews the foundational work underpinning our system, while Section~\ref{sec:system} details its architecture and key components. Section~\ref{sec:exp} provides insights and evaluation findings from the system's implementation, discussing its implications for data professionals and the evolution of BI systems. Section~\ref{sec:relatedWork} reviews related work on query generation. Finally, Section~\ref{sec:threats} discusses threats to validity while Section~\ref{sec:future} concludes with findings and directions for future research. 

\section{Running Example}
\label{sec:example}
The core aim of \AutoBIR is to enable direct and intuitive interaction between users (data analysts, subject matter experts, executives, etc.) and data systems using natural language, thus formalizing analytics requirements with greater precision and efficiency than with existing traditional approaches. This example demonstrates several core functionalities: (1)~\textit{the initial business question}, (2)~\textit{an automatically generated analytics query}, (3)~\textit{a natural language interpretation of the query}, (4)~\textit{the requisite data model in an ontological format}, (5)~\textit{mappings to physical data sources}, and (6)~\textit{actual test cases produced as part of the requirements specification process}.

\begin{figure*}[htbp]
    \centering
    \begin{minipage}{\textwidth}
        \centering
        \includegraphics[width=\linewidth]{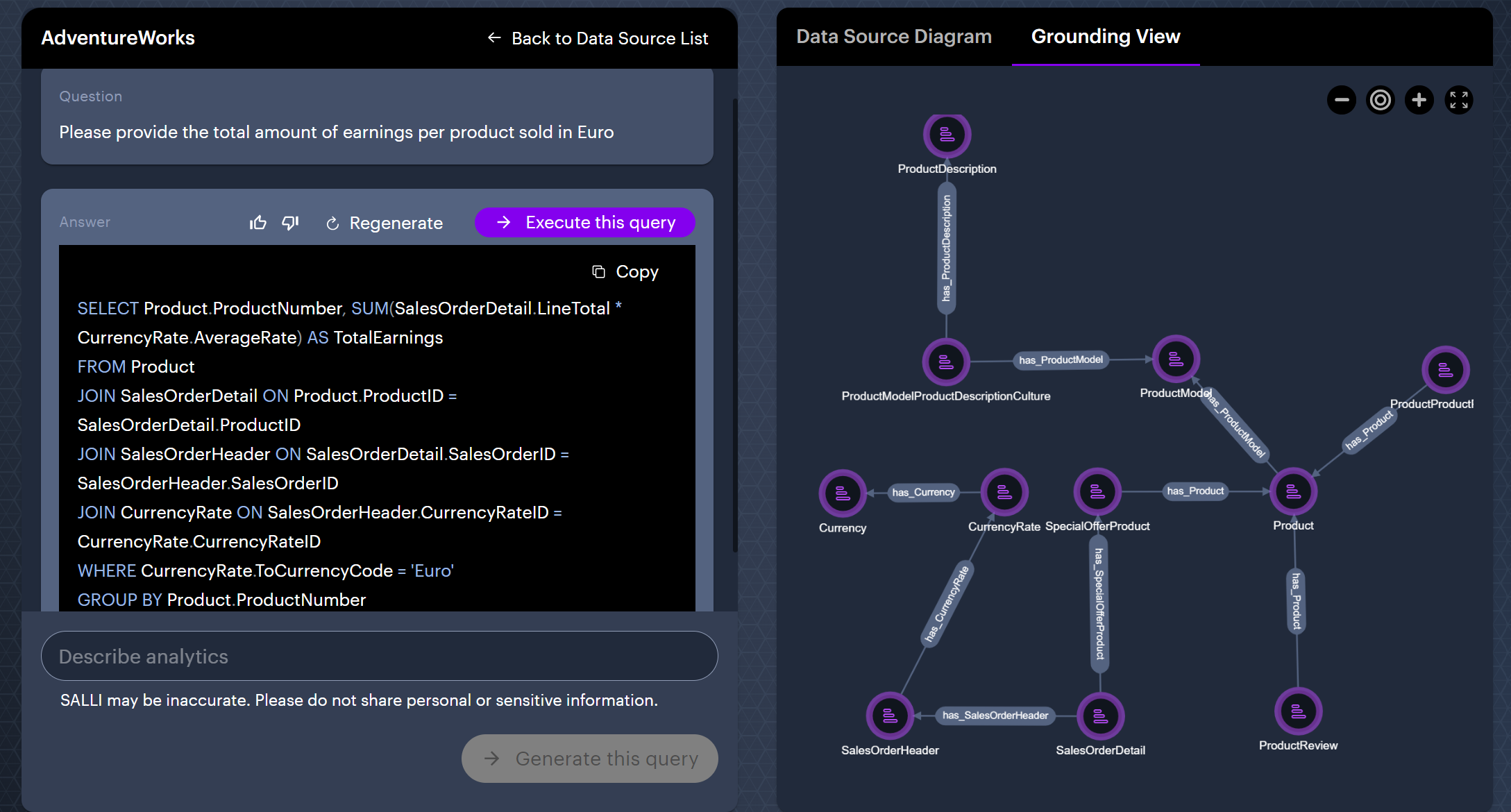}
        \caption{A view of the \AutoBIR system interface including the user's business question (left, top), the query generated by our system (left, bottom), and the ``grounding'' sub-ontology provided as context for the Generative AI (right).\\}
        \label{fig:system_main}
    \end{minipage} \hfill
    
    \begin{minipage}{\textwidth}
        \centering
        \includegraphics[width=\linewidth]{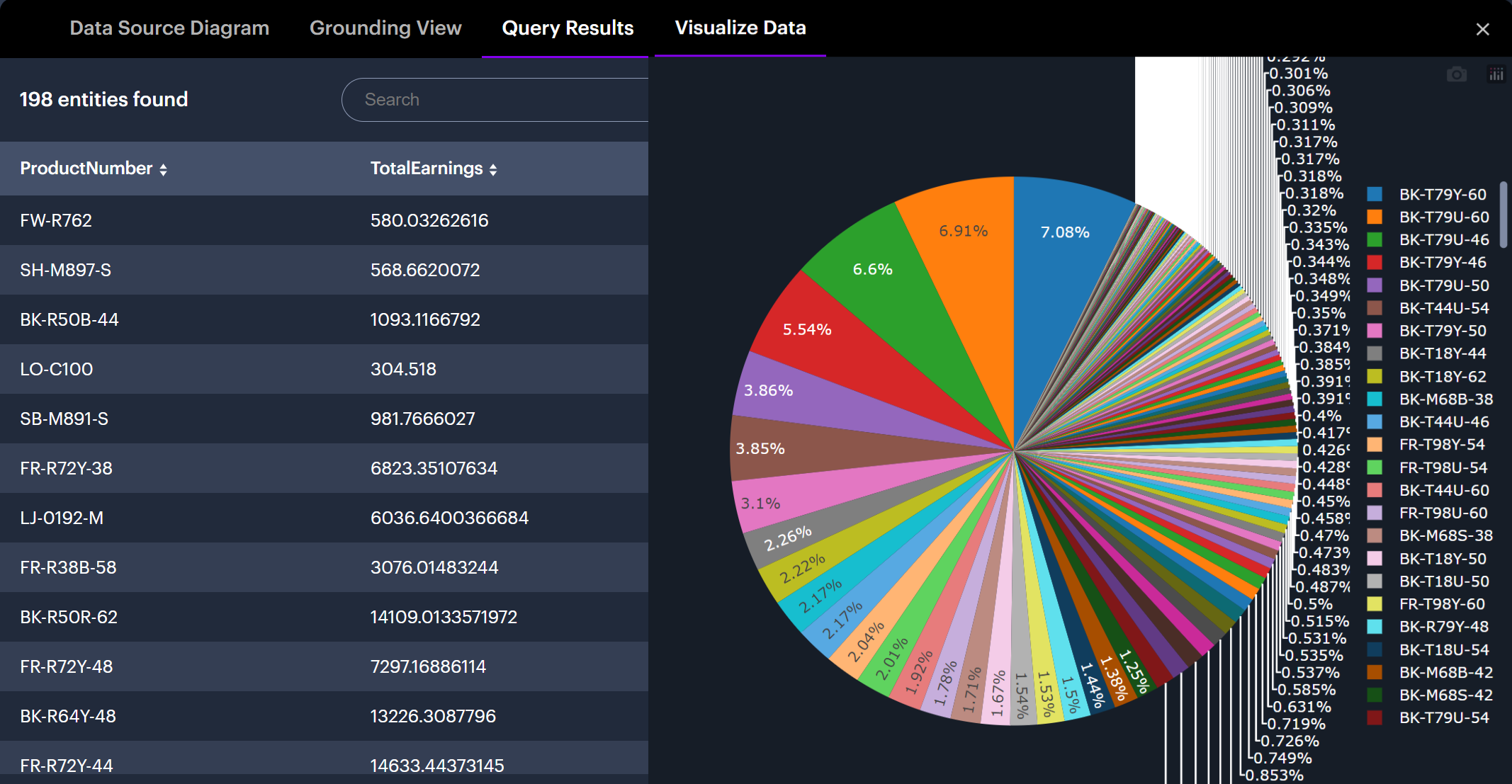}
        \caption{A view of the \AutoBIR system interface including query execution results (left) and a data visualization generated using Generative AI (right).}
        \label{fig:vis}
    \end{minipage}
\end{figure*}

To showcase these capabilities, we use the open-source database developed by Microsoft: \texttt{AdventureWorks2014}~\cite{microsoft_adventureworks_2012}. This database serves as a benchmark for SQL Server database design. It simulates the information system of a bicycle store, encompassing 71 tables and hundreds of columns, providing a rich dataset for diverse analytical queries. In the example, a user asks for the total amount of products sold in Euro, and an SQL query and explanations of the query, along with a graph-based visualization of a sub-ontology to show relevant datasets provided (see graph nodes in the \textit{Grounding View} of Fig.~\ref{fig:system_main}). The user then executes the query directly on the \texttt{AdventureWorks2014} database, receiving the report displayed in Fig.~\ref{fig:vis} (left). In this example, the term ``Earnings'' was employed in the business question, rather than technical terms like ``Sales'', ``Sales Amount'', or any reference to \texttt{CurrencyRate} or \texttt{CurrencyCode} in the query tables. 

This indicates that the tool uses semantic search to interpret the intent based on the data model. The requirements specifications generated from this process include the initial business questions, the SQL query for analytics. Table~\ref{tbl:example} shows a snippet incorporating the business question, example SQL query (Listing~\ref{lst:query}), query results (Figure~\ref{fig:vis}, left), supporting data model (Listing~\ref{lst:data_model}), and bindings to the database tables (Listing~\ref{lst:binding}). Importantly, the data model presented to the user abstracts physical database structures into conceptual relationships, for example, showing foreign-key relationships as object properties rather than explicit joins, enhancing usability for non-technical stakeholders. The four joins between tables \texttt{Salesorderdetails}, \texttt{Salesorderheader}, \texttt{CurrencyRate}, and \texttt{Currency} in the example query in Fig.~\ref{fig:system_main} appear as a path between the corresponding entities in the data model (see the nodes with the same labels in the \textit{Grounding View}).

The system leverages Generative AI (GenAI) to explain query logic and produce reports in formats that go beyond traditional tables offering visual aids that deepen user comprehension. Figure~\ref{fig:vis} illustrates this capability, with the generated reports depicted on the left and their visualizations on the right. These reports provide a robust foundation for validating and refining BI requirements. This innovative application of GenAI demonstrates how advanced AI technologies can facilitate more streamlined and scalable interactions with data-intensive systems, enabling users to navigate and refine their BI requirements with greater accuracy and clarity.

\begin{table}[htbp]
\centering
\caption{System and User interaction example}
\label{tbl:example}
\begin{tabular}{|p{0.9\linewidth}|}
\hline
\textbf{Business Question (User)}: ``Please provide the total amount of earnings per product sold in Euro''.  
\begin{lstlisting}[language=SQL, caption=Query Generation Response (System), label=lst:query]
SELECT Product.ProductNumber, SUM(SalesOrderDetail.LineTotal*CurrencyRate.AverageRate) AS TotalEarnings
FROM Product
JOIN SalesOrderDetail ON SalesOrderDetail.ProductID = Product.ProductID
JOIN SalesOrderHeader ON SalesOrderDetail.SalesOrderID = SalesOrderHeader.SalesOrderID
...
WHERE CurrencyRate.ToCurrencyCode = "Euro"
GROUP BY Product.ProductNumber
\end{lstlisting}
\\ \hline
\textbf{Query Explanation Response (System)}: 
The results display the total earnings of each product, calculated by multiplying the sales line total by average currency exchange rate, specifically for transaction converted to Euros, with the data organized by product names.

\\ \hline
\textbf{Execution Results and Test Case (System)}: 
 The query execution results are displayed in the table in See Figure~\ref{fig:vis} (left). 
 The table includes the two columns \texttt{ProductNumber} and \texttt{TotalEarnings}, and total earnings per product. 

 \begin{lstlisting}[caption=Ontology Snippet (System), label=lst:data_model]
 @Class@ specialofferproduct {
     @Data Property@: rowguid VARCHAR, 
     @Data Property@: ModifiedDate VARCHAR, 
     @Data Property@: SpecialOfferID INT, 
     @Data Property@: ProductID INT, 
     @Object Property@: has_product REFERENCES product
}
@Class@ productdocument { 
     @Data Property@: ModifiedDate VARCHAR, 
     @Data Property@: ProductID INT, 
     @Data Property@: DocumentNode INT, 
     @Object Property@: has_product REFERENCES product 
} ...
\end{lstlisting}
 \begin{lstlisting}[caption=Bindings Snippet (System), label=lst:binding]
 @Class@ c: specialofferproduct => @Table@ t: specialofferproduct, @Table@ t2: product
    c.rowguid => t.rowguid,
    c.SpecialOfferID=> t.SpecialOfferID,
    c. ProductID =>t. ProductID,
    c.ModifiedDate=>t.ModifiedDate ,
    c.has_specialoffer=>t.specialofferproduct,
    c.has_product=> t2.specialofferproduct
    ...
\end{lstlisting}
 \\ \hline
\end{tabular}
\end{table}




\hspace{1em}
\section{Preliminaries}
\label{sec:prem}
These are the component technologies utilized in \AutoBIR:

\begin{itemize}
\setlength{\leftmargin}{2em} 

\item \textbf{OWL}. The Web Ontology Language (OWL) is a robust,
expressive language specifically designed for the formulation
knowledge representation~\cite{w3c_owl_working_group_owl_2012}. With its ability to provide formal and structured vocabularies, OWL is an effective tool for representing various domains of interest. It facilitates the specification of classes, properties, and their instances, thus enabling comprehensive descriptions of concepts and the relationships between them. In \AutoBIR, we utilize OWL as a medium to abstractly represent the logical models of heterogeneous data sources defined by diverse formalism, such as relational databases~\cite{codd_relational_1970} and JSON schemas~\cite{bray_javascript_2014}. The primary means of exchanging OWL is through the Resource Description Framework (RDF)~\cite{w3c_rdf_2014}. In this context, ``ontology'' represents the logical data model, while ``physical model'' refers to the representations in the data source system. 

\item \textbf{Data Bindings}. Data bindings serve as formal languages utilized to articulate bespoke mappings between physical data models and RDF datasets. In \AutoBIR, OWL ontologies represent data sources, while bindings like R2RML~\cite{w3c_r2rml_2012} are harnessed to express
the mappings between these physical data models and their
respective ontologies. These bindings enable seamless switching between ontologies and the physical models, supporting query generation and system interactions with the underlying information systems, e.g. Databricks~\cite{databricks_inc_databricks_2022}, Microsoft SQL\cite{microsoft_sql_2022}, and Stardog\cite{stardog_union_stardog_2022}.

\item \textbf{Semantic Search}. Our research integrates semantic search capabilities using vector databases (Vector DBs) such as Milvus~\cite{zilliz_milvus_2022} and Pinecone~\cite{pinecone_systems_inc_pinecone_2022}. These databases excel in identifying items that are most ``similar'' to a given search query by leveraging embeddings, i.e., fixed-length numerical vectors, that represent items in a high-dimensional space to find similar items through k-Nearest Neighbor algorithms~\cite{arya_algorithms_1993}. Furthermore, contemporary Vector DBs combine vector-based and traditional filtering techniques, utilizing Approximate Nearest Neighbor~\cite{muja_scalable_2014} and support hybrid search~\cite{jing_when_2024} for efficiency. In \AutoBIR, semantic search matches user queries to ontological concepts in a knowledge graph, creating a sub-ontology that generates SQL queries via an LLM. 

\item \textbf{Text2Query}. Recent research has investigated generating queries from text~\cite{li_can_2023,yu_spider_2018}. The conventional problem formulation adopted in this study infers a mapping from a textual description of a query, referred to as a \textit{business question} in natural language, into \textit{SQL queries}~\cite{noauthor_isoiec_2016} that most closely aligns with it. Furthermore, in recent years, researchers have proposed methods for other query languages, such as Cypher~\cite{neo4j_inc_cypher_2022} and SPARQL~\cite{w3c_sparql_2013}. 

\end{itemize}

\section{\AutoBIR System Overview}
\label{sec:system}
Although {\AutoBIR} is currently utilized internally as part of Accenture's service offerings and is not yet publicly available, this section describes its core components. These core components include \textit{Setup} tools for connecting to new data sources and \textit{Run-time} tools that process user business questions.

\begin{figure*}[t]
\centerline{\includegraphics[width=\textwidth]{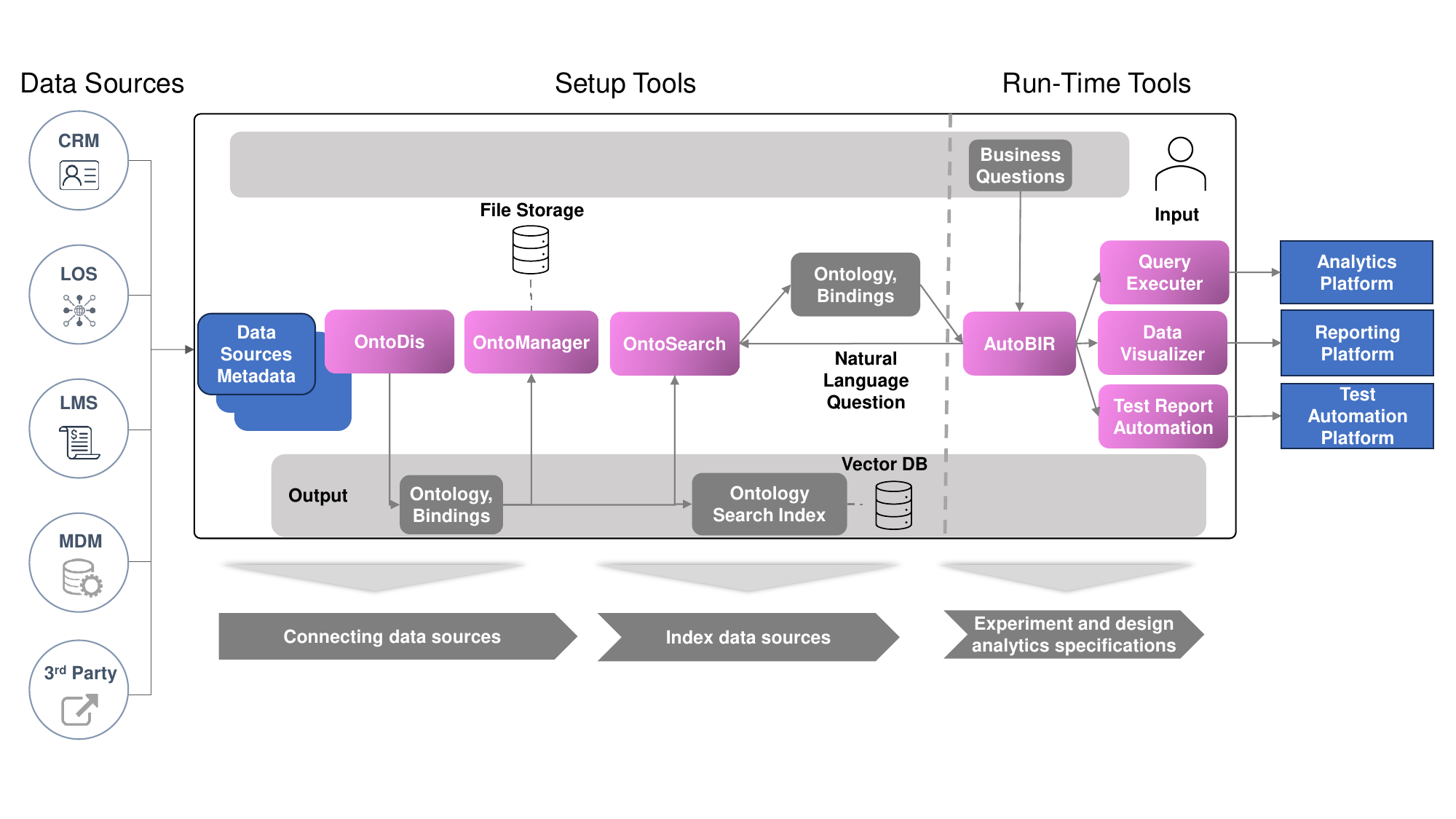}}
\caption{High-level system overview: To upload a new data source, the system invokes the \textit{Setup} tools (left of the dashed line). To support the human operator in defining analytics specification, the system uses the \textit{Run-time} tools (right of the dashed line). When a user asks a business question, the system triggers the \AutoBIR tool to generate a query and an explanation. The system uses additional tools to execute the query, generate visualizations, and archive the results as test cases.}
\label{fig:sys_arch}
\end{figure*}

\textbf{Setup Tools}: Figure~\ref{fig:sys_arch} illustrates the system's high-level architecture and the integration sequence for a new data source. The process begins with \textit{OntoDis}, the Ontology Discovery tool, which constructs an ontological data model from the source system metadata and maps it to the physical model.  Next, \textit{OntoManager}, a tool for cataloging, catalogs details like connection parameters, the derived data model, and associated bindings. Finally, \textit{OntoSearch} is an ontology search tool designed to enrich a vector database with a semantic search index tailored for the new ontology. It enables dynamic interpretation of natural language queries and highlighting pertinent entities within the ontology at \textit{Run-Time}.    

\textbf{Run-time Tools}: When a user asks a business question, \AutoBIR generates a corresponding executable query. The  \textit{Query Executor} then runs the query and retrieves the results while the \textit{Data Visualizer} offers graphical representations of the data insights concurrently. The \textit{Test Report Automation} tool archives these results for subsequent testing endeavors. To improve explainability and align with user intent, an optional feature generates a natural language explanation of the query, providing feedback on how the request was interpreted.

\subsection{OntoDis: Generating an Ontology and Data Bindings}
\label{sec:onto_dis}
The system enables business users to define analytics requirements across diverse and distributed data sources, such as traditional databases and external data services. It abstracts physical data models into an ontological model, standardizing data representations under OWL. This approach supports selective integration, shifts focus from physical to conceptual models, and sidestepping constraints like 3NF~\cite{codd_further_1972} efficiency. It also allows merging data from different sources by reconciling similar concepts, such as `Customer' with `Client'.

The \textit{OntoDis} tool is essentially using metadata from data sources to create an OWL ontology with mappings to the physical data model. It preserves these bindings by applying R2RML, \textit{OntoDis} conventions. Through iterative refinements, \textit{OntoDis} abstracts and conceptualizes the data model, resulting in a comprehensive OWL ontology and its bindings.

\textbf{Model Refinement}: \textit{OntoDis} enhances the ontology through targeted model edits, such as class deletion, renaming, partitioning, merging, and property removal. It refines the model based on policies combining conditions and modification actions, adjusting entities by evaluating their labels and relationships.

\textbf{Model Refinement}: \textit{OntoDis} enhances the ontology through targeted model edits like class deletion, renaming, partitioning, merging, and property removal. It refines the model based on policies-driven conditions and modification actions, adjusting entities by assessing their labels and relationships.

\begin{mdframed}[backgroundcolor=gray!20, linewidth=0pt]
\textbf{Example}. Consider a refinement policy for a class with the suffix \texttt{Properties}, linked to a class with the same name minus the suffix. The refinement operation, ``collapse entity to referencing class(es)'', removes this entity and merges its properties with the related class. For example, in an ontology with classes \texttt{File} and \texttt{FileProperties}, linked by an object property, this policy would remove \texttt{FileProperties}. All properties previously associated with \texttt{FileProperties} are then directly attributed to \texttt{File}, resulting in an updated ontology.
\end{mdframed}

\textit{OntoDis} integrates customizable model refinement policies to enhance data comprehension and improve LLM predictive accuracy for natural language queries. For example, it renames mislabeled constructs (including tables and columns) with cryptic abbreviations or foreign languages transliterations such as ``saptaah\_ka\_din'' for ``day\_of\_the\_week'' in Hindi, or ``Yīzhōu\_zhōng\_de\_tiān'' in Chinese. These policies advance data modeling, especially in extensive, dynamic systems, reduce manual intervention, and increase adaptability, hence enriching the semantic model beyond its physical counterpart.

\subsection{OntoManager: Data Source and Ontology Management}
\label{sec:onto_manager}
\textit{OntoManager} catalogs data models, bindings, and access permissions according to data governance policies, using a multi-tenant strategy to manage user responsibilities. It supports data-source versioning and collections management, ensuring efficient data governance and access control to meet diverse user needs and operational requirements.

\subsection{OntoSearch: Searching an Ontology Catalog}
\label{sec:onto_search}
To navigate extensive ontology catalogs, \textit{OntoSearch} is used to create semantic search indices for new ontologies. It processes each entity using labels and annotations to construct string representations. It optionally also includes associated properties for classes. A pre-trained or fine-tuned language model transform these strings into embeddings stored in a vector database. \textit{OntoSearch} also enhances entity indices with metadata attributes such as the origin data source or the entity type for hybrid searches. For example, identifying the top \textit{k} classes related to the word ``Helmet'' in the \texttt{AdventureWorks2014} database might reveal “Product” as a semantically relevant table, illustrating the system's capacity to interpret ``Helmet'' as a product category. The precision of search outcomes and query formulation is greatly influenced by the string representations, prompting \textit{OntoSearch} to offer customizable settings for selecting which annotations to include in the embedding generation, like prioritizing class definition annotations or ignoring version annotations.

\textbf{Searching for a Sub-Ontology}: To translate a user's query, a sub-ontology with relevant entities is identified. The process begins by extracting nouns from the query to locate corresponding classes and their related properties. The search expands by exploring the shortest paths in the ontology's graph until reaching a predefined boundary. Concept pairs are prioritized for expansion based on the descending cosine similarity of their embeddings, hence promoting a focused and comprehensive mapping to the user's intent.

\textbf{Remark}: The proposed algorithm recognizes that queries often traverse multiple entities in the data model, in a way similar to SQL joins. Since users may not fully know these entities and relationships, the algorithm uses semantic searches to identify relevant entities from user questions. The goal is to extract a sub-ontology that includes all necessary entities for accurate query prediction (recall) while minimizing extraneous elements (precision).  

\subsection{\AutoBIR: Query Generation and Self-Debugging}
\label{sec:AutoBIR}
The essential function of the system is to convert business inquiries into operational queries through a Text-to-Query transformation. This research area, with a particular focus on Text-to-SQL that has seen significant growth and activity~\cite{katsogiannis-meimarakis_survey_2023,qin_survey_2022}, aims to articulate business needs accurately. Enhancements include: 1)~query generation and customization for large systems; 2)~user feedback integration and conversational continuity; 3)~self-debugging and error correction; 4)~explaining query logic (Query-to-Text), and 5)~enabling cross-platform queries across distributed systems.

\textbf{Query Generation and Customizations}: 
To translate a user query (business question in Fig.~\ref{fig:sys_arch}, \textit{OntoSearch} first identifies a relevant sub-ontology. An LLM then crafts a query in the desired query language, guided by this sub-ontology and predefined query generation rules. Since the identified ontology operates independently from the physical data model, we utilize its bindings for mapping to the actual data source. The query execution strategy varies by data source. For example, an SQL query for an SQL Server database references relational tables, while for a knowledge graph platform like Stardog~\cite{stardog_union_stardog_2022}, which runs SPARQL queries and supports OWL ontologies, the LLM generates SPARQL queries using terms from the OWL sub-ontology.

At system activation, the query language-specific guidelines, example question-query pairs, and a method for translating OWL constructs and bindings into the selected query language are provided. It is assumed that the sub-ontology and its bindings cover the necessary information for querying the physical data, as supported by experiences with platforms such as Stardog~\cite{stardog_union_stardog_2022} and Databricks~\cite{databricks_inc_databricks_2022}.


\textbf{Conversation History and User Feedback}: The system enhances  Text-to-Query frameworks by maintaining a dynamic dialogue with the user, preserving the last \textit{k} messages to keep the conversation context. The context informs follow-up questions posed to the LLM, improving the relevance and precision of the response and elevating the quality of conversational engagement.

\begin{mdframed}[backgroundcolor=gray!20, linewidth=0pt]
\textbf{Example}. Imagine a user querying our system about the database featured in our running example (Sect.~\ref{sec:example}) with: ``what kind of helmets are in stock?''. The \AutoBIR tool uses \textit{OntoSearch} to identify the sub-ontology pertinent to the inquiry. It then formulates a prompt that includes instructions for SQL query generation, a string depiction of the relevant data model, and the user's original question. Listing~\ref{lst:prompt_example} shows an excerpt from that prompt, followed by an LLM response shown in Listing~\ref{lst:query_results}.
\end{mdframed}

\begin{lstlisting}[label=lst:prompt_example, caption=Example LLM prompt generated by the system.]
Task: Given a data schema, *and* a free-text question, produce an SQL query that matches it.
- Guideline: ...
- Guideline: ...
...

Your query must refer to the following schemas:

CREATE TABLE Product (
  ProductID int,
  Name varchar,
  ProductNumber varchar,
  PRIMARY KEY (ProductID)
);
 
CREATE TABLE ProductInventory (
  Shelf varchar,
  Bin int,
  ProductID int,
  LocationID int,
  PRIMARY KEY (ProductID, LocationID),
  FOREIGN KEY (ProductID) REFERENCES Product(ProductID)
);
...
 
Question: what kinds of *helmets* *are* *in* stock?
\end{lstlisting}
\begin{lstlisting}[label=lst:query_results, caption=Example response from the LLM.]
Query:
SELECT Name FROM Product WHERE Name LIKE '%helmet%' AND ProductID IN (SELECT ProductID FROM ProductInventory WHERE Quantity > 0)
\end{lstlisting}

\textbf{Self-Debugging and Error Correction}: The system features a self-debugging capability that uses three types of \textit{checkers} (syntax, semantic, and execution-based) to debug and fix potential query issues. The \textit{syntax checker} ensures proper query grammar (e.g., proper SQL), the \textit{semantic checker} verifies consistency with the ontology's entities, types, and taxonomies, and checks alignment between the user's question and the generated query. The \textit{execution-based checker} runs the query through an analytics engine to detect execution errors or exceptions.

If any \textit{checker} detects an issue, the system feeds this information into the next prompt cycle, guiding the LLM to correct previous mistakes and generate an accurate query. This iterative approach improves query generation, streamlining the transformation of user questions into executable queries. In our experiment, the checkers ensured the correctness of query conditions rather than validating query results. Deterministic checkers, like the \textit{syntax checker}, maintain safety by rejecting invalid queries while non-deterministic checkers, like the \textit{semantic checker}, may occasionally reject valid queries and thus require careful design and evaluation to minimize inaccuracies.

\begin{mdframed}[backgroundcolor=gray!20, linewidth=0pt]
\textbf{Example}. Listing~\ref{lst:bad_query} includes an example of an invalid query generated by the LLM. In this scenario, the \textit{semantic checker} tests the query and outputs the error messages in Listing~\ref{lst:checkers_results}. The system parses the errors and produces the error message in Listing~\ref{lst:error_correction}. Subsequently, the next call to the LLM includes those findings in order to avoid these errors.
\end{mdframed}
\begin{lstlisting}[caption=An erroneous query (System)., label=lst:bad_query]
SELECT FirstName, LastName, Shift    
FROM BadTableName   
WHERE Department = 'Quality Assurance' 
\end{lstlisting}
\begin{lstlisting}[caption=Checkers' results (System)., label=lst:checkers_results]
Status: 'Invalid'  
Checker Type: 'Semantic'
Error message: 'Table BadTableName does not exist' 
\end{lstlisting}
\begin{lstlisting}[caption={Instruction for subsequent prompt (System).},label=lst:error_correction]
Generated query may be invalid because:
- *Table* BadTableName does *not* exist. 
Only generate queries with the provided tables.
\end{lstlisting}

\textbf{Explainability}: When a query is generated in a formal language like SQL, an explanation is provided for non-expert users. Technical users can review SQL statements on the physical model, while business users can examine the query's purpose through an explainable statement. A prompt is created for the LLM to explain the query to a non-expert in a chosen style: \textit{Compact}, \textit{Verbose}, \textit{Formal}, \textit{Simple}, or \textit{Precise}. The prompt includes the original question and the sub-ontology used for query generation as additional context. 

\textbf{Querying Distributed Data Sources}: Traditional query generation typically targets single data sources. By leveraging knowledge graph platforms like Stardog~\cite{stardog_union_stardog_2022}, our approach extends to multiple data sources represented as virtual knowledge graphs. This enables \textit{OntoSearch} to traverse and locate sub-ontologies across various ontologies, enabling comprehensive searches. A SPARQL query is then generated and executed across the knowledge graph, accessing the distributed data repository.

\textbf{Remark}: Integrating data sources requires reconciling similar concepts by aligning them to a common ontology, refining the initial model generated by \textit{OntoDis}. Future research will explore additional alignment techniques.


\subsection{Generating Reports and Data Visualizations}
\label{sec:data_vis}
After generating an executable query, the user can run it on the source database. The system executes the query using the \textit{Query Executer} in Fig.~\ref{fig:sys_arch}, on an analytics platform that either includes a data source or is a data platform (Fig.~\ref{fig:sys_arch}, \textit{Analytics Platform}). \AutoBIR retrieves the result sets with optional pagination, and displays the results on a dashboard. If the user opts to visualize the data, an LLM is prompted to generate graphical visualizations of the results in the \textit{Data Visualizer} in Fig.~\ref{fig:sys_arch}. For this purpose, we include guidelines on supported visualizations in a connected graphic library or tool (Fig.~\ref{fig:sys_arch}, \textit{Reporting Platform}), a small subset of results, for example, the first 10 rows and columns, guidelines on the expected format, and a few graphics generation code examples. The expected format includes the code for producing the selected visualizations.   The process repeats if errors occur, halting upon successful execution or after a set number of iterations. The system currently uses  \textit{plotly.express}~\cite{plotly_technologies_inc_plotly_2022}, a Python library, as the visualization format, which can be configured by modifying the associated graphic tool or library and provided examples.

Finally, the generated reports and visualizations effectively communicate desired outcomes and can be archived for future references or uploaded as test cases to the test report automation platform (see Fig.~\ref{fig:sys_arch}, \textit{Test Report Automation} component). This is done based on analytical specifications and a snapshot of the data used to produce these reports. This ensures that each visualization meets immediate analytical needs and contributes to a repository of insights for future validation and review.

\section{Evaluation and Lessons Learned}
\label{sec:exp}

This section presents insights and feedback from implementing and refining the \AutoBIR system across four domains: Security, Air Defense, Retail, and Banking. Each domain featured unique databases and objectives, and interactions were reviewed by 23 Subject Matter Experts (SMEs) from five countries and four organizations. These SMEs, which included data professionals (analysts, engineers, and scientists), served as \textit{key informants}, providing feedback for our qualitative evaluation~\cite{Minichiello2008}. Structured sessions were used to engage the key informants to test interactions. In line with the Design Science Research methodology~\cite{Hevner2004}, feedback received, which included enhancements to the framework and identification of domain-relevant use cases, was used in iterations to modify the framework. This evaluation approach provided practical insights into the toolset's adaptability and effectiveness across various contexts. Lessons learned are detailed next.

\subsection{Enhancing Data Analysis: Interactive Feedback and User-Centric Design}
\label{subsec:usr_feedback}
The interactive feedback mechanism connecting user questions with generated queries and supporting data models serves several key purposes. It provides real-time insights into existing data capabilities, helps users refine their queries for accuracy, and offers opportunities to improve ontologies, enhancing data comprehension for both users and the LLM. 

The precision of generated queries and underlying data models is crucial for the success of no-code platforms, especially for business users and data professionals.  The system's integration of LLMs with an interactive feedback loop enhances user-friendliness and efficiency in crafting analytical specifications. Additional features include the ability to regenerate queries, manually adjust outputted queries, inquire about the data model and schema, and edit data visualizations. These enhancements aim to match user intent with fewer modifications. 

A significant use case emerged from collaborating with data professionals using the toolset for the semantic interpretation and deeper understanding of datasets. Their objective was to expedite data analysis and swiftly prototype analytical dashboards, demonstrating the toolset's value in streamlining complex data analysis and enhancing data accessibility and insights generation.

\subsection{Technical System Requirements}
\label{subsec:tch_eval}
Our exploration targeted the diverse utility of LLMs, especially in directing user and SME queries towards data model grounding. 
Discussions with SMEs led to the development of several functionalities detailed in Sect.~\ref{sec:AutoBIR}.  
Frequent requests included support for various databases and storage types, as well as the ability to generate queries in languages beyond SQL over distributed data sources. 

With a semantic layer of ontologies abstracting connected databases, our system uses bindings to perform model composition and fusion by generating multiple queries. This enables a modular and flexible architecture that can transition across different query languages and BI engines.  

\subsection{On the Importance of a Good Data Model}
\label{subsec:data_model}
Data Engineers have long recognized the importance of a \textit{logical} data model to complement the \textit{physical} data model~\cite{kuper_logical_1993}. However, logical models are often missing, and physical models are filled with abbreviations, cryptic and ambiguous names, or phonetic transcriptions of non-English terms, making them hard to understand and limiting query effectiveness. The pivotal role of LLMs in \AutoBIR{}-like systems highlights the need for semantic layers that LLMs can easily interpret. 

This challenge opens new research avenues focused on deducing, evaluating, and refining formal ontological representations and establishing best practices for their design and maintenance. Developing logically-sound and semantically-rich models is crucial for improving the interface between human cognition and machine understanding, enhancing data interaction.
As companies transition into the era of GenAI applications, neglecting to refine physical data models or failing to implement logical models may hinder their ability to adopt tools like \AutoBIR. 

The suite of tools introduced in this study demonstrates the potential to create a searchable catalog of semantic data representations with a natural language interface. This provides a user-friendly summary of possible data integrations, enhancing decision-making processes. These advancements position businesses to leverage GenAI technologies, ensuring their data infrastructure is compatible with and optimized for these innovations.

\subsection{Security Considerations in LLM-Driven Query Generation}
\label{subsec:security}
Using an LLM for query generation introduces several security concerns~\cite{liu_refining_2023}. There is a risk of generating queries that could unintentionally modify information systems, disclose unauthorized data, or lead to SQL injection vulnerabilities if not properly configured~\cite{pedro_prompt_2023}. Additionally, fine-tuning the LLM with database data might inadvertently reveal stored values, risking sensitive information leakage that bypasses data source-level access controls. If a malicious actor compromises the LLM service responsible for query generation~\cite{yang_stealthy_2024}, it could produce harmful outputs, from computationally burdensome to subtly incorrect, or data-exposing. 

Entities using or developing such systems must meticulously manage the data fed into the LLM, implement protective measures, establish security best practices, and limit access to Personally Identifiable Information (PPI) or sensitive data during the fine-tuning and prompting stages.

\subsection{Implications for Future BI Systems Design and Usage}
\label{subsec:future}
Our recent exploration investigated \AutoBIR{}'s potential to transform BI systems by democratizing BI activation for business users through discussions with data domain executives. The modular architecture of \AutoBIR facilitates rapid integration with various data sources, remote querying of federated systems, and reduces knowledge barriers. This promotes the democratization of personal BI dashboard prototypes, easing the burden on data engineering teams and improving productivity in dashboard design. The toolset also enhances BI systems design by improving analytics discoverability, cataloging analytics by descriptions or shared data, and optimizing data structures for common analytics. 

These advancements are critical for developing data products and forming data mesh grids~\cite{dehghani_data_2022}, adding a new dimension to data product development and organization. Subject Matter Experts, serving as key informants, believe \AutoBIR can revolutionize reporting and analytics design, reducing labor costs and shortening delivery times. A large-scale user study is planned to evaluate the system's effectiveness in improving specification tasks.
  
\section{Related Work in Query Generation Research}
\label{sec:relatedWork}

\textbf{Query Generation}: Text-to-SQL leverages sequence-to-sequence models with various encoding methods, as summarized by Katsogiannis and Koutrika~\cite{katsogiannis-meimarakis_survey_2023}, including bidirectional LSTM~\cite{shi_incsql_2018}, CNNs~\cite{choi_ryansql_2021}, BERT~\cite{hwang_comprehensive_2019}, and graph neural networks~\cite{bogin_representing_2019}. Intermediate representations from Rubin et al.~\cite{rubin_smbop_2020} aid in natural language to SQL transformations. 

Decoding methods split into sketch-based slot-filling, which uses fixed templates and struggles with flexibility, and generative methods that require extensive data~\cite{choi_ryansql_2021}. Sketch-based models~\cite{hui_improving_2021,hwang_comprehensive_2019} are template-dependent,  while generative models~\cite{rubin_smbop_2020} demand substantial training resources. Liu et al.~\cite{liu_refining_2023} investigate zero-shot prompting with LLMs for Text-to-SQL using the Spider dataset~\cite{yu_spider_2018}. Their approach employs prompts for table comprehension, logical reasoning, and table-to-text transformation, achieving strong results with minimal examples. This study underscores LLMs' zero-shot potential, showing their adaptability to NLP tasks without extensive training.

Recent research also emphasizes the need to address critical challenges in Text-to-SQL systems when applied to real-world scenarios. Studies have shown that execution accuracy declines significantly on real-world datasets compared to academic benchmarks like Spider, with contributing factors including long context lengths, ambiguous question formulations, and complex nested queries~\cite{Ganti2024-TextToSQL-Failures}. To mitigate these challenges, \AutoBIR leverages ontology-based abstraction and semantic search to reduce schema complexity, while its interactive feedback loop allows iterative query refinement, improving robustness and performance in real-world applications.

Moreover, advancements in Text-to-SQL approaches, such as PURPLE~\cite{Ren2024-LLM-SQL}, demonstrate the value of schema pruning and good selection of logical composition operators to improve query generation. PURPLE enhances logical operator composition through tailored demonstrations, guiding LLMs in managing complex SQL structures~\cite{Ren2024-LLM-SQL}. 

\textbf{Ensuring Query Integrity and Automated Debugging}: Query accuracy is essential in automated generation. Chen et al.\cite{chen_teaching_2023} employ LLMs to evaluate query validity based on user input and query content. Our approach extends this by incorporating a semantic checker for ontology interaction and syntax/execution checkers for deterministic validations. Additionally, self-debugging draws from error-handling strategies by Saben et al.\cite{saben_enabling_2024} and Yasunaga et al.~\cite{yasunaga_break-it-fix-it_2021}, aiming to refine error resolution. Our work highlights the importance of dialogue-based functionalities to enhance query validation and interpretation.

\textbf{Advancements in Query Evaluation}: Text-to-SQL research has advanced through benchmarks like the Spider dataset~\cite{yu_spider_2018}, which improved query generation and inspired leaderboard-focused studies~\cite{li_graphix-t5_2023,wang_rat-sql_2019}. Due to Spider's limitations with single-domain schemas, the BIRD dataset~\cite{li_can_2023} was introduced, covering 37 domains with a competitive leaderboard, while WikiSQL remains widely used for single-table queries~\cite{zhong_seq2sql_2017}. Standard evaluation metrics—Component Matching, Exact Matching, and Execution Accuracy—are commonly used, with Li et al.~\cite{li_graphix-t5_2023} introducing the Valid Efficiency Score for added depth. \AutoBIR evaluation leverages Spider and BIRD datasets, focusing on Exact Matching and Execution Accuracy. While BIRD exposes weaknesses in existing methods, it lacks comprehensive testing for dialogue systems. CoSQL~\cite{yu_cosql_2019} addresses this by simulating user-database interaction, revealing the importance of enhancing benchmarks to account for dialogue-based system capabilities. Key criteria for such systems include maintaining conversational context, adapting to user feedback, and handling complex queries. 

The recent release of Spider 2.0~\cite{lei2024spider20evaluatinglanguage} introduces a new level of complexity in Text-to-SQL research, providing a benchmark specifically designed for real-world enterprise workflows. With databases that often exceed 1,000 columns, stored across platforms like BigQuery and Snowflake, Spider 2.0 requires models to process extensive contexts, including metadata, dialect documentation, and project-level codebases, while generating multi-step queries of over 100 lines. These challenges go beyond traditional metrics, underscoring the need for robust and scalable models that integrate reasoning with diverse SQL dialects and workflows.

In the context of \AutoBIR, Spider 2.0's emphasis on enterprise scenarios aligns closely with the system's goal of automating BI requirement elicitation in complex data environments. Spider 2.0 highlights the gap between existing Text-to-SQL models and the needs of enterprise-level systems, with initial evaluations reporting a stark performance drop (17\% task completion on Spider 2.0 compared to 91.2\% on Spider 1.0). This reflects the additional challenges of translating high-level business queries into SQL workflows while considering dynamic contexts and user-driven refinements. 

Addressing these challenges represents a significant step toward enabling systems like \AutoBIR to manage real-world BI systems' intricate requirements, demonstrating generative AI's broader potential in advancing data-driven decision-making processes.

\section{Threats to Validity}
\label{sec:threats}
This section discusses important categories of threats to the validity of our proposed framework. These are internal, external, and construct validity~\cite{Verdecchia2023-ThreatsToValidity}. Mitigating these threats is important to improve the reliability and generalizability of our findings.

\subsection{Internal Validity}
\label{subsec: Internal Validity}
Internal validity refers to the degree to which a study accurately establishes a causal relationship between its variables, free from the influence of confounding factors~\cite{Sculley2018WinnersCO}. In the context of research with LLMs, internal validity ensures that observed improvements in model performance are genuinely due to changes in the model or training approach, not external factors like data leakage, biased datasets, or flawed evaluation methods. 

Threats to the internal validity of our work and how we mitigate them include: 

\begin{itemize}
\setlength{\leftmargin}{2em} 

\item \textbf{Bias in Data Selection}. A potential limitation is the use of the \texttt{AdventureWorks2014} database~\cite{microsoft_adventureworks_2012}, which may not fully represent the diversity of real-world Business Intelligence systems. To mitigate this bias, we incorporated ontologies and query benchmarks, such as the Spider~\cite{yu_spider_2018} and BIRD datasets~\cite{li_can_2023}. These datasets cover diverse domains and structures, enabling a broader evaluation of \AutoBIR{}'s performance. Future iterations will expand testing to include additional heterogeneous datasets such as the new (and much larger) Spider 2.0~\cite{lei2024spider20evaluatinglanguage} dataset. 

\item \textbf{Quality of Generated Ontologies}. The quality of the ontologies generated by the \textit{OntoDis} tool is critical to the accuracy of semantic search and query generation in \AutoBIR. Errors such as incorrect mappings or missing relationships could have occurred, propagating through the system and impacting the accuracy of semantic search and query generation. To address this, we incorporated iterative refinement mechanisms. These measures help ensure logical consistency and alignment with physical data schemas. 

\item \textbf{User Feedback Integration}. Reliance on user feedback from \textit{key informants} for the iterative improvements of \AutoBIR could have introduced potential biases. These could have been due to varying expertise levels among the informants, or to tendencies to provide overly positive feedback. To address this, \AutoBIR includes a feedback standardization mechanism that translates user inputs into structured refinements for ontologies and queries. Additionally, by enabling a conversational interface that maintains dialogue history, the system improves user understanding and minimizes repetitive feedback errors. Future work will explore more robust usability studies with users. 

\item \textbf{Generative Model Dependency}. Errors in LLM outputs, such as inaccurate SQL queries, could compromise the system's results. To mitigate this, \AutoBIR integrates self-debugging mechanisms (see Section~\ref{sec:AutoBIR}) that employ syntax, semantic, and execution-based validation. Additionally, an explainability module provides users with a natural language interpretation of generated queries, enabling iterative refinement and reducing ambiguity. 
\end{itemize}

\subsection{External Validity}
\label{subsec: External Validity}
External validity refers to the extent to which the findings of a study or the performance of a model can be generalized beyond the specific conditions of the research, such as different datasets, environments, or user contexts~\cite{Sculley2018WinnersCO}. It ensures that the results observed under controlled experimental conditions are applicable and reliable in real-world scenarios. For LLMs, external validity evaluates whether model behaviors, such as accuracy or coherence demonstrated in benchmark tests, are consistent across diverse applications and unseen tasks.

Threats to the external validity of our work and how we mitigate them include: 

\begin{itemize}
\setlength{\leftmargin}{2em} 

\item \textbf{Scalability and Generalization}. While \AutoBIR has been tested on small-to-medium-scale datasets, its scalability to enterprise-grade environments remains to be explored. Challenges in scaling similar systems have been documented in enterprise-grade evaluations such as Spider 2.0~\cite{lei2024spider20evaluatinglanguage}. To mitigate this, we designed a modular architecture to support distributed data querying and scalable vector databases, such as Milvus~\cite{zilliz_milvus_2022} and Pinecone~\cite{pinecone_systems_inc_pinecone_2022} for semantic search. Initial scalability tests with federated data sources demonstrated promising results, but further evaluation in larger environments is planned.

\item \textbf{Applicability to Non-Standardized Data}. \AutoBIR reliance on pre-defined schemas and OWL ontologies limits its applicability to semi-structured or unstructured data. In order to mitigate this threat, we included dynamic schema adaptation features within \textit{OntoDis}, enabling schema inference for new data sources. 

\end{itemize}

\subsection{Construct Validity}
\label{subsec: Construct Validity}
Construct validity refers to the degree to which a study or model accurately measures or represents the theoretical concept it is intended to assess. For LLMs, it ensures that the evaluation metrics or tasks genuinely reflect the underlying capabilities being studied, such as reasoning, coherence, or factuality, rather than unrelated artifacts or superficial patterns in the data.

Threats to the construct validity of our work and how we mitigate them include: 

\begin{itemize}
\setlength{\leftmargin}{2em} 

\item \textbf{Evaluation Metrics}. While query accuracy and execution correctness are standard metrics, they do not fully capture user satisfaction or the practical utility of \AutoBIR. To address this, we included qualitative, informal evaluations from 23 Subject Matter Experts (SMEs) who served as \textit{key informants} across multiple domains. These evaluations provided insights into usability and task alignment, complementing quantitative measures.

\item \textbf{Comparisons with Existing Systems}. \AutoBIR{}'s performance was benchmarked against state-of-the-art Text-to-SQL and semantic search systems using datasets like Spider~\cite{yu_spider_2018} and BIRD~\cite{li_can_2023}, which simulate real-world enterprise scenarios. These comparisons highlighted \AutoBIR{}'s competitive accuracy and superior ability to handle user feedback in iterative query generation. However, comparisons with other technologies for accelerating BI dashboard creation, especially beyond Test-to-SQL (e.g., regarding explanations and selected visual representations), remain to be performed.

\end{itemize}

\section{Conclusions and Future Work}
\label{sec:future}

This paper contributes a no-code framework, \AutoBIR, designed to streamline the creation of functional analytics requirements in Business Intelligence (BI) environments. Through a conversational interface, the system interprets natural language inputs to generate query code, data models, explanations, execution outcomes, and visual representations of data. This adaptable tool supports extensive customization, such as incorporating new query languages, modifying data model discovery processes, and leveraging alternate platforms for query execution to suit specific BI needs. Business users and data professionals can use simple prompting mechanisms to tailor responses and refine visualizations to meet their unique requirements.

The framework was implemented and refined during client engagements, addressing diverse objectives, including envisioning future BI systems, enhancing data products and integration workflows, and prototyping new analytics solutions. It simplifies technical complexities for business users by generating business-focused reports while providing detailed technical specifications for data engineers to actualize solutions. By bridging these perspectives, the system will help accelerate analytics development and ensure alignment between technical and business goals.

As development continues, the framework aims to democratize data analytics further, making it more accessible and streamlined for users. Its no-code automation capability marks a significant advancement, enabling users to define analytics requirements and explore new possibilities with minimal technical expertise. Continuous improvements will integrate cutting-edge research and emerging technologies to enhance performance and user experience, addressing present demands while anticipating future challenges.

The mitigation measures implemented to address the many threats to validity described in the previous section help ensure that \AutoBIR is robust, scalable, and user-friendly. Future work will build on these mitigations by expanding the size and diversity of datasets, refining ontology generation methods, and conducting large-scale user studies and additional comparisons.

This work also emphasizes the need for ongoing research into advanced techniques such as automatic ontology discovery, enhanced semantic search, and robust model grounding. It highlights the importance of developing methodologies for guided, secure, and explainable query generation, alongside establishing benchmarks for effectiveness, usability, and security in data analytics. The goal is to provide a reliable, user-friendly solution that meets evolving standards and unlocks new opportunities in the BI domain.

\bibliographystyle{IEEEtranDOI}  
\bibliography{refs} 

\end{document}